\documentclass[sn-mathphys-num]{sn-jnl}

\usepackage{amsmath,amsfonts,bbm,mathrsfs,amssymb,ulem}
\usepackage{color}  
\definecolor{violet}{rgb}{0.561,0.0,1}
\usepackage{bm} 
\usepackage{hyperref}
\hypersetup{colorlinks=true,urlcolor=blue,anchorcolor=blue,citecolor=blue,filecolor=blue,
            linkcolor=blue,menucolor=blue, linktocpage=true,pdfproducer=medialab}
\usepackage[english]{babel}

\newcommand*{\INFNFR}{Istituto Nazionale di Fisica Nucleare, Laboratori Nazionali di Frascati, C.P. 13, 00044 Frascati, Italy}
\newcommand*{\LNGS}{Istituto Nazionale di Fisica Nucleare, Laboratori Nazionali del Gran Sasso, Assergi, 67100, Italy}
\newcommand*{\NICPB}{Laboratory of High Energy and Computational Physics, NICPB, R\"avala 10, 10143, Tallin, Estonia}
\newcommand*{\LAPTh}{LAPTh, Université Savoie Mont-Blanc et CNRS, 74941 Annecy, France}

\def\equationautorefname~#1\null{Eq.\,(#1)\null}

\def\eq#1{{Eq.~(\ref{#1})}}

\newcommand{\be}{\begin{equation}}
\newcommand{\ee}{\end{equation}}
\newcommand{\ba} {\begin{equation}\begin{aligned}}
\newcommand{\ea} {\end{aligned}\end{equation}}
\newcommand{\bg} {\begin{equation}\begin{gathered}}
\newcommand{\eg} {\end{gathered}\end{equation}}

\newcommand{\mX}{m_{X_{17}}} 
\newcommand{\RBe}{R_{\rm Be}}

\newcommand{\MeV}{\ \text{MeV}}

\newcommand{\bea}{\begin{eqnarray}}
\newcommand{\eea}{\end{eqnarray}}

\title{Combined Evidence for the $X_{17}$ Boson After PADME Results on Resonant Production in Positron Annihilation}

\author[a]{\fnm{Fernando} \sur{Arias-Aragón}}\email{fernando.ariasaragon@lnf.infn.it}
\author[b]{\fnm{Giovanni} \sur{Grilli di Cortona}}\email{giovanni.grilli@lngs.infn.it}
\author[a,c]{\fnm{Enrico} \sur{Nardi}}\email{enrico.nardi@lnf.infn.it}
\author[d]{\fnm{Claudio} \sur{Toni}}\email{claudio.toni@lapth.cnrs.fr}
\affil[a]{\INFNFR}
\affil[b]{\LNGS}
\affil[c]{\NICPB}
\affil[d]{\LAPTh}

\abstract{The Positron Annihilation into Dark Matter 
Experiment at the Laboratori Nazionali di Frascati has reported an excess of $e^+e^-$ final-state events from positron annihilation on fixed-target atomic electrons. 
While the global significance remains at the $(1.77\pm 0.15)\,\sigma$ level, the excess is centered around $\sqrt{s} \sim 17\,\text{MeV}$, coinciding with the invariant mass at which anomalous $e^+e^-$ pair production has previously been observed in nuclear transitions from excited to ground states in $^8$Be, $^4$He and  $^{12}$C, thereby strengthening the case for a common underlying origin, possibly involving a hypothetical new $X_{17}$ boson.
We discuss the significance of this independent accelerator-based evidence. Combining it with existing nuclear physics results, we obtain a value for the $X_{17}$ mass of $\mX = 16.88 \pm 0.05\,\text{MeV}$, reducing the uncertainty from nuclear physics determinations by more than a factor of two, and mitigating the impact of poorly known correlations among their systematic errors.}

\begin{document}
\maketitle
\flushbottom

\section{Introduction}
{\it A spectre is haunting particle physics -- the spectre of the $X_{17}$ boson}~\cite{Marx2002-MARCM}.
About ten years ago, a puzzling anomaly was observed by the ATOMKI collaboration 
in the angular correlation spectrum of $e^+e^-$ pairs emitted in $^8$Be$^*$(18.15\,MeV)
transition to the ground state~\cite{Krasznahorkay:2015iga}. 
The enhancement of the $e^+e^-$ signal observed at large angles relative to internal pair creation (IPC) exhibited a significance of approximately $6.8\,\sigma$, and could be interpreted as the emission of a neutral isoscalar boson with a mass $\mX \sim 17\,\text{MeV}$. This intriguing possibility promptly attracted the attention of particle theorists, who began exploring the nature and properties of the hypothetical $X_{17}$ boson~\cite{Feng:2016jff}. At the same time, 
it was  quickly recognized that the most unambiguous way to verify the existence of the $X_{17}$  
was through a dedicated particle physics experiment designed to observe the resonant production of the new boson via positron annihilation on a fixed target~\cite{NewDirections}. Such an approach would be free from nuclear physics uncertainties and entirely independent of potential systematic effects affecting the original nuclear physics experiment. 
A proposal to utilize the positron beam from the Beam Test Facility (BTF) linac at the Laboratori Nazionali di Frascati (LNF), impinging on a thick $^{74}$W target to produce the new boson via the resonant process $e^+e^- \to X_{17}$, in combination with  the detector of the  Positron Annihilation into Dark Matter Experiment (PADME) to search for 
electron/positron pairs from $X_{17}\to e^+e^-$ decay, was first discussed 
in~\cite{NewDirections}.  
Subsequently, a detailed study of the sensitivity reach of a PADME-like experimental setup was published in~Ref.~\cite{Nardi:2018cxi}.

In the meantime, after an upgrade of  the experimental apparatus, the $^8$Be anomaly was confirmed~\cite{Krasznahorkay:2018snd}, and 
during the following years,  other anomalies were  repeatedly observed 
by the ATOMKI collaboration in the $e^+e^-$ angular correlation spectra 
from de-excitation of  
$^4$He~\cite{Krasznahorkay:2019lgi,Krasznahorkay:2019lyl,Krasznahorkay:2021joi}, 
$^{12}$C~\cite{Krasznahorkay:2022pxs}  and 
 of the Giant Dipole Resonance (GDR) of $^8$Be~\cite{Krasznahorkay:2023sax,Krasznahorky:2024adr}.
An anomaly consistent with the ATOMKI findings was recently 
observed   also  
in an experiment conducted at VNU-University of Science, that investigated IPC in the de-excitation of $^8$Be$^*$(18.15\,MeV)~\cite{Anh:2024req}.
Remarkably, the kinematic features of 
all these observations appear  to be consistent with 
a single explanation in terms of a new boson with mass 
around 17\,MeV emitted in the transition of the 
excited nuclei to the ground 
state~\cite{Feng:2016jff,Feng:2016ysn,Feng:2020mbt,Barducci:2022lqd}.
Another experiment aiming to verify the existence of the X17 particle
was conducted at the Paul Scherrer Institute 
with the MEG II (Muon Electron Gamma) detector~\cite{MEGII:2024urz}, using 
the same $^7$Li(p,$e^+e^-$)$^8$Be  reaction used 
by ATOMKI. In this case, no significant evidence of the $X_{17}$  particle was found.
However, since MEG II reports results that are compatible with the ATOMKI ones within $1.5\,\sigma$, they disfavour but do not rule out the $^8$Be anomaly. Instead, their measurement can be consistently incorporated into a global fit, as was already suggested in  Ref.~\cite{Barducci:2025hpg}.
Additionally, results from the SINDRUM-I experiment~\cite{Eichler:2021efn} set rather stringent constraints on the hypothetical $X_{17}$~\cite{Hostert:2023tkg,DiLuzio:2025ojt}. However, the corresponding limits strongly depend on the spin and parity of the particle, as well as on its couplings to quarks. Establishing a proper limit would thus require a dedicated analysis of the underlying dynamics, a task  beyond the scope of this work.
In the near future, other experimental collaborations 
will provide additional results on the hypothetical $X_{17}$, see e.g. Refs.~\cite{Azuelos:2022nbu,Cortez:2023ycv,Gustavino:2024wgb,Gongora-Servin:2025uda,LDMA2025LNL}.
An interesting overview of published and forthcoming experimental results is presented in Ref.~\cite{Krasznahorky:2024adr},
while a review of possible theoretical interpretations 
can be found in Ref.~\cite{Alves:2023ree}.

The strategy ultimately adopted for $X_{17}$ searches at LNF using the PADME detector differs from the initial proposal in Ref.~\cite{Nardi:2018cxi}, as the experiment employed the existing thin (100$\,\mu$m) polycrystalline active diamond target instead of a thick tungsten target. 
The main aspects  of the $X_{17}$ resonant production 
with this new experimental setup were presented in Ref.~\cite{Darme:2022zfw}, which also provided preliminary sensitivity estimates based on the approximation that the target atomic electrons are free and at rest.
This is an important point since, as already emphasized in Ref.~\cite{Nardi:2018cxi}, even for a low $Z$ material such as carbon, the momentum distribution of bound atomic electrons 
significantly broadens the distribution of
 centre-of-mass (c.m.) energies, by a factor a few times larger 
that a typical intrinsic  beam energy spread of $O(1\%)$. 
However, shortly after the PADME collaboration completed data-taking in the fall of 2022, a comprehensive study of atomic electron velocity effects on positron annihilation on fixed targets was also completed~\cite{Arias-Aragon:2024qji}. The corresponding corrections have been fully incorporated into the PADME experimental analysis, as outlined in Ref.~\cite{Bertelli:2025mil}, where  the PADME data analysis strategy is discussed in detail.
The results from the unblinding of the PADME $X_{17}$ data sample were first presented in a talk at the LDMA workshop in Genoa~\cite{LDMA2025},  shortly thereafter in a general seminar at LNF~\cite{LNFGenSem2025}, and were subsequently published in Ref.~\cite{PADME:2025dla}.  Intriguingly, the data show an excess 
of $e^+e^-$ events peaked at $\sqrt{s} = 16.90\,$MeV. 
Although the significance is moderate ($(1.77\pm0.15)\,\sigma$ global, $2.5\,\sigma$ local) the signal presents some thought-provoking features: 
\begin{itemize} \itemsep -2pt
\item As we will demonstrate, the location of the excess is in excellent agreement with the value derived from a global fit to the available nuclear physics data.

\item{The excess appears across more than one adjacent bin, consistently with the expected broadening of the c.m. energy due to beam energy spread and atomic electron motion.}

\item{Two quantitatively similar excesses at the same c.m. energy were observed in both scans performed by the PADME collaboration, separated by a time interval of 1.5 months.}
\end{itemize} 

In the next section, we will examine the impact of the PADME measurement on the determination of the $X_{17}$ mass, and on drastically reducing the dependence of the associated confidence regions
on the poorly known correlations among the systematic errors of the nuclear physics results. As we will demonstrate, despite the moderate significance of the excess, the remarkably small uncertainty in the c.m. energy of the annihilation process (which we expect will remain the dominant error 
in the $X_{17}$ mass determination) enables a reduction in the uncertainty on the $X_{17}$ mass by more than a factor of two.
On the other hand, after including the PADME result, the difference between the confidence regions estimated under the two extreme assumptions of fully correlated versus uncorrelated systematic errors for the ATOMKI measurements, becomes negligible. Thus, our analysis provides precise and robust prior information, which will be of key importance for future searches for the $X_{17}$ and 
related data analyses.

\noindent
\section{Global fit to the $X_{17}$ data}
Defining a consistent set of data from published results is not straightforward, because the full information on systematic errors
is not always provided. We take the results for the first ATOMKI $^8$Be measurement~\cite{Krasznahorkay:2018snd} from the reanalysis 
presented in Ref.~\cite{Krasznahorkay:2015iga}, where 
a result for the partial widths ratio 
\be
\label{eq:RBe}
R_{\rm Be} \equiv 
\frac{\Gamma(^8{\rm Be}^*\to ^8{\rm Be} +X)}{\Gamma(^8{\rm Be}^*\to ^8{\rm Be}+ \gamma)} {\rm Br}(X\to e^+e^-) 
\ee
is also reported ($R_{\rm Be} = (6.8\pm 1.0) \times 10^{-6}$). 
Since in Ref.~\cite{Krasznahorkay:2018snd} a systematic error on 
the mass determination 
is not provided, we assign to this measurement $\sigma^{\rm syst} =0.50\,$MeV as in 
the original paper~\cite{Krasznahorkay:2015iga}.
We take  from Ref.~\cite{Krasznahorkay:2018snd} also 
the data for the second  ATOMKI $^8$Be measurement. 
This measurement was performed after the upgrade of the experimental apparatus, so we assign to the mass determination a systematic error 
$\sigma^{\rm syst} =0.20\,$MeV in accordance with the 
$^4$He and $^{12}$C measurements that were performed with a similar  experimental setup. 
The  value for the partial width ratio for this measurement 
quoted in~\cite{Krasznahorkay:2018snd} is $R_{\rm Be} = 
(4.7\pm 2.1) \times 10^{-6}$. Note that for both values 
of $R_{\rm Be} $ only the statistical error is reported. 
Since our primary focus is on the impact of the PADME measurement on the determination of $\mX$, omitting a systematic uncertainty for the two $R_{\rm Be}$ measurements does not affect our conclusions.

\begin{table*}[t!!]
    \centering
    \renewcommand{\arraystretch}{1.5}
    \begin{tabular}{|c|c|c|c|}
    \hline
    Nucleus (MeV)  & $\mX\,$(MeV) & Experiment & Ref. \\
    \hline
    $^8$Be$^*$(18.15)  &  $16.86\pm0.06\pm0.50$ & Atomki &  \cite{Krasznahorkay:2015iga,Krasznahorkay:2018snd} \\
    \hline
        $^8$Be$^*$(18.15)  &  $17.17\pm0.07\pm0.20$ & Atomki &  ~\cite{Krasznahorkay:2018snd} \\   
    \hline
    $^4$He$^*$(20.21/21.01) & $16.94 \pm 0.12 \pm 0.21 $ & Atomki & \cite{Krasznahorkay:2021joi} \\
    \hline
   $^{12}$C$^*$(17.23) & $17.03 \pm 0.11 \pm 0.20 $ & Atomki & \cite{Krasznahorkay:2022pxs} \\
    \hline
 $^{8}$Be$^*$(GDR) & $16.95 \pm 0.48 \pm 0.35 $ & Atomki & \cite{Krasznahorkay:2023sax,Krasznahorky:2024adr} \\
    \hline
     $^{8}$Be$^*$(18.15) & $16.66 \pm 0.47 \pm 0.35 $ & VNU-UoS & \cite{Anh:2024req} \\
    \hline
         $^{8}$Be$^*$(17.64/18.15) & $\qquad < 16.81 $   [$R_{\rm Be} = 6\cdot 10^{-6}$] $\qquad$ & MEG II & \cite{MEGII:2024urz} \\
    \hline
      $e^+e^- \to X_{17}$ & $16.90\pm 0.02\pm 0.05$ & PADME & \cite{PADME:2024pwg,Bossi:2025ptv} \\
    \hline
    \end{tabular}
\vspace{6 pt}
    \caption{The $X_{17}$ mass measurements used in our fit. In the second column, 
    for the nuclear physics results the first uncertainty  is the statistical error, the second is the systematic. The central values and the statistical errors in the first two rows are 
    taken from Table~1 of Ref.~\cite{Krasznahorkay:2018snd}, while the systematic error in the first measurement 
    is taken from Ref.~\cite{Krasznahorkay:2015iga}. For the second measurement,  
    no systematic error was reported in Ref~\cite{Krasznahorkay:2018snd}. This measurement was performed
    with an improved experimental setup, which was subsequently used  for the $^4$He and $^{12}$C measurements
    shown in the third and fourth rows.  We therefore assign to this measurement the same systematic error reported for the $^4$He and $^{12}$C data, which stems mainly from the uncertainty in the beam position.
    For the MEG II measurement we give the $90\%$ CL upper limit 
    on $\mX$ corresponding to the best fit point $R_{\rm Be}= 6\cdot 10^{-6}$.  
    For the particle physics result in the last row the first uncertainty comes from the beam energy step (0.75\,MeV) and   the second from the absolute precision in the beam momentum (1.5\,MeV)~\cite{PADME:2024pwg}.
    The statistical uncertainty from the mass fit is subdominant and has been neglected (see footnote~\ref{foot-pvalues}).
    }
    \label{tab:summary}
\end{table*}

Significant correlations are expected among the various $\mX$ measurements performed with the ATOMKI 
experimental setup. 
Unfortunately, the experimental papers lack detailed 
information on these correlations, complicating the combination of the different results.  
However, both Refs.~\cite{Krasznahorkay:2015iga}  
and~\cite{Krasznahorkay:2018snd} mention that the leading 
systematic error is caused by the instability of the beam position 
on the target. In Ref.~\cite{Krasznahorkay:2021joi} systematic errors were estimated by performing  Monte Carlo (MC) 
simulations, confirming that the uncertainty on the precise location of the beam spot on target represents the main source of  systematic errors. 
In Refs.~\cite{Krasznahorkay:2022pxs} and 
\cite{Krasznahorky:2024adr} the same MC simulations described  
in~\cite{Krasznahorkay:2021joi}  were used to assign a systematic uncertainty to the results (no systematic error is given in Ref.~\cite{Krasznahorkay:2023sax}).  
Finally, the VNU-UoS experiment in Ha Noi~\cite{Anh:2024req} used the same Li$_2$O targets provided by ATOMKI,\footnote{A. Krasznahorkay, private communication.} and similarly identified the main source of uncertainty as arising from the imprecision in the beam-spot position.

\begin{figure*}[t!]
    \centering
    \includegraphics[width=0.49 \textwidth]{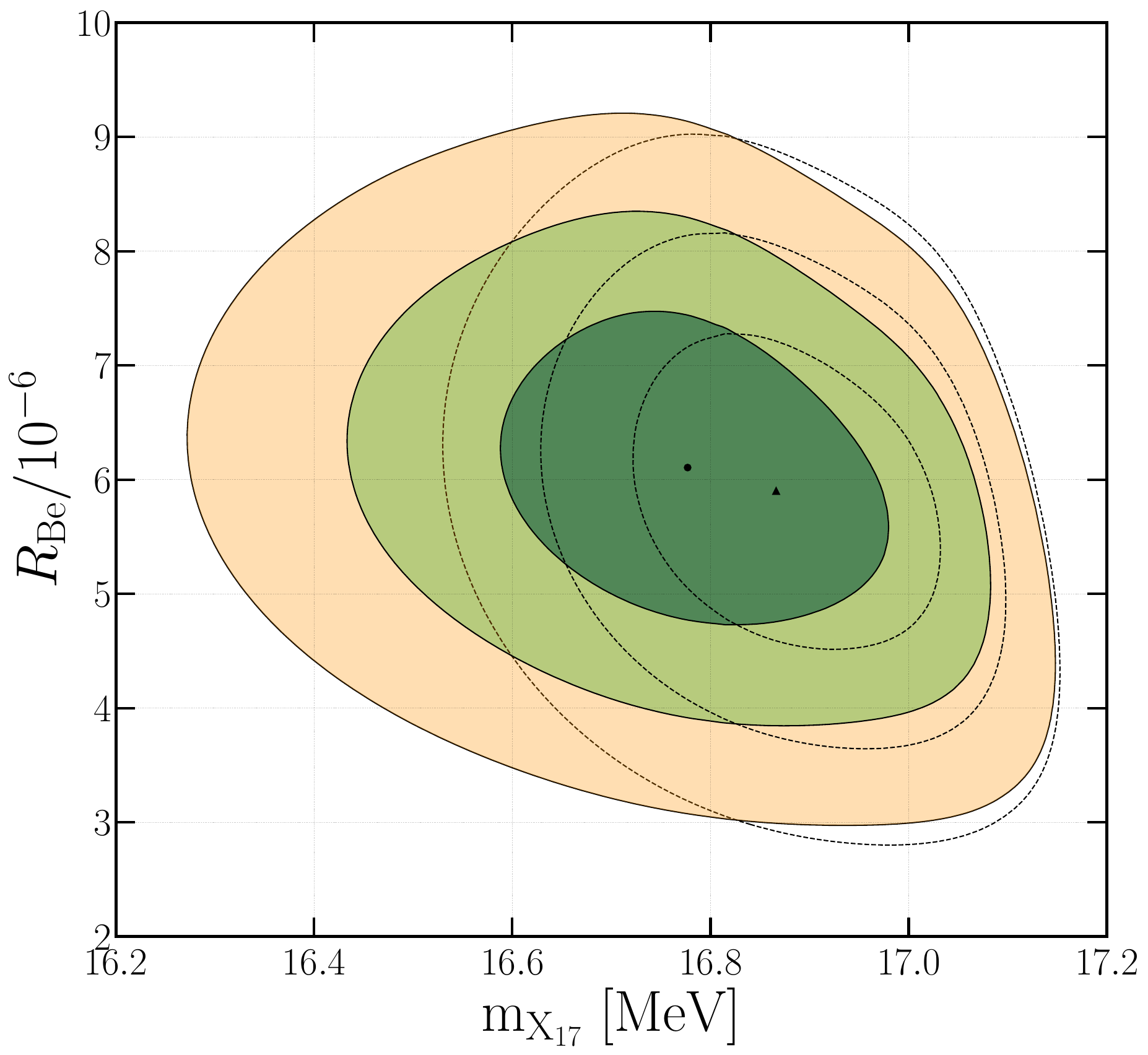}
    \includegraphics[width=0.49 \textwidth]{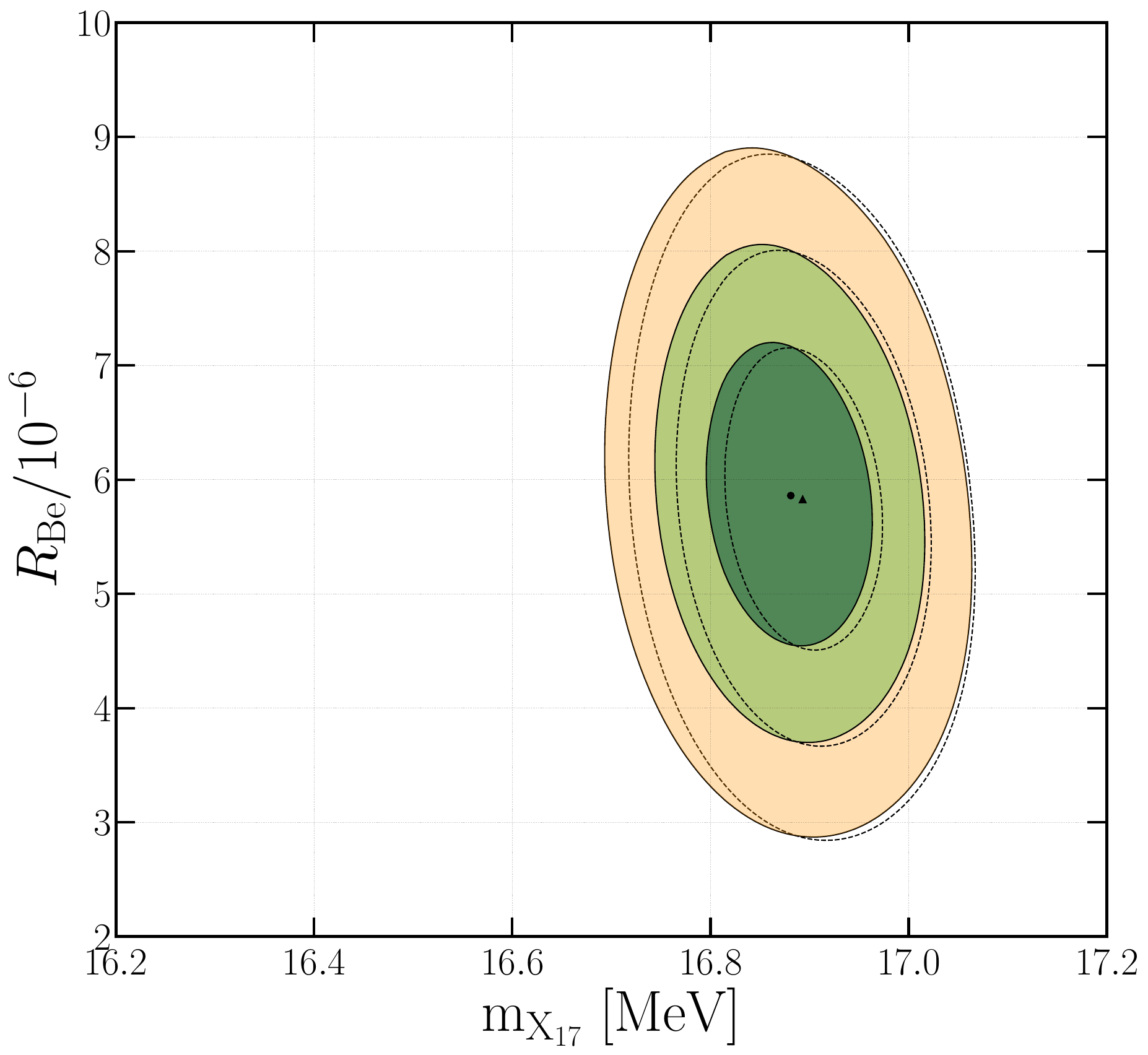}
    \caption{
Fits to the $X_{17}$ mass. Left: without the PADME result. Right: with the PADME result included.  The different contours correspond to the $1\sigma$, $2\sigma$ and $3\sigma$ regions of a $\Delta\chi^2=\chi^2-\chi^2_\mathrm{min}$ variable with $2$ degrees of freedom.
The solid  contours and the black circles represent 
the confidence regions and best fit point obtained 
by assuming full correlation among systematic uncertainties in the nuclear physics results (see text). 
The dotted black contours and the triangle markers correspond to the opposite assumption
of uncorrelated systematic uncertainties.
    }
    \label{fig:Fits}
\end{figure*}

\begin{figure}[t!]
    \centering
    \includegraphics[width=0.7 \textwidth,trim={1pt 1pt 1pt 1pt}, clip]{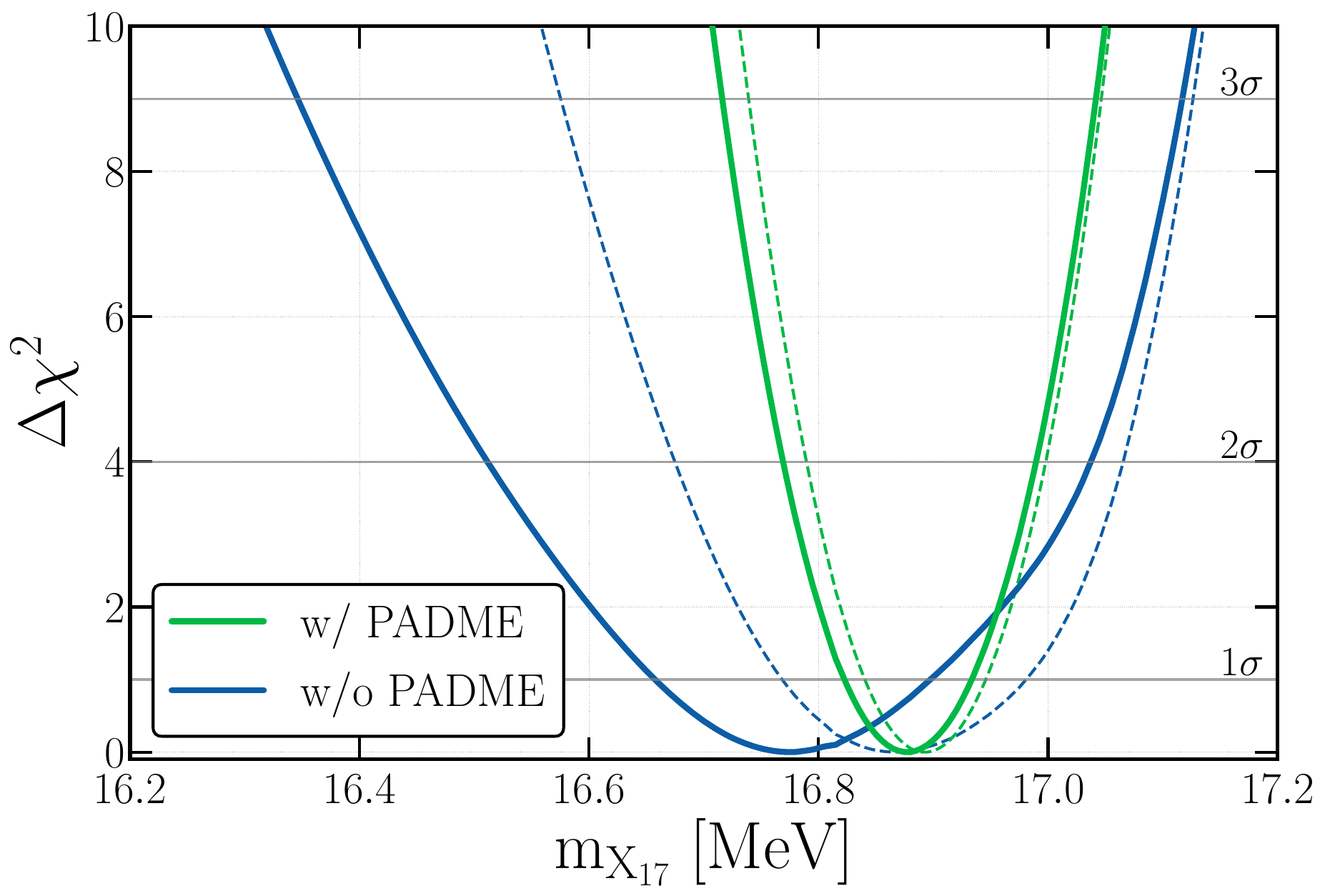}
    \caption{
Value of the $\Delta\chi^2=\chi^2-\chi^2_\mathrm{min}$ for the $X_{17}$ mass marginalized over $R_{\rm Be}$. Green (blue) lines are with (without) the inclusion of the PADME measurement. 
Continuous (dashed) lines correspond to assuming fully correlated (uncorrelated) 
 systematic uncertainties for the nuclear physics results.
The gray horizontal lines corresponds to the $1\sigma$, $2\sigma$ and $3\sigma$ 
of a $\chi^2$ variable with one degree of freedom. 
    }
    \label{fig:Chi2}
\end{figure}

More in detail, the systematic errors from the 
uncertainty in the beam position on the target quoted in Refs.~\cite{Krasznahorkay:2015iga,Krasznahorkay:2021joi,Krasznahorkay:2022pxs,
Krasznahorky:2024adr,Anh:2024req} are respectively   
$[\pm 0.50,\,\pm 0.21,\,\pm 0.20,\,\pm 0.35,\,\pm 0.35]\,$MeV, and   
account for the {\it entirety} of the systematic errors reported   
in the corresponding final results (see Table~\ref{tab:summary}).
Ref.~\cite{Krasznahorkay:2018snd} states that the discrepancy in the $X_{17}$ particle mass fitted from the two data 
sets that they have analysed  (respectively $m_{X_{17} }=  16.86(6)$\,MeV 
and $m_{X_{17} }= 17.17(7)\,$MeV)
is likely due to the unstable  beam position,  without giving, however, an  explicit number. 
As mentioned above,  we then assign  a systematic error of 0.20 MeV in accordance 
with the $^4$He and $^{12}$C measurements.
This number is also compatible with the size of the systematic  uncertainty required 
to reconcile the mentioned discrepancy.
This body of information, albeit not completely exhaustive, confirms the common origin of the leading systematic uncertainties in the various measurements and allows one to conclude that a strong correlation exists among the  errors of the ATOMKI and VNU-UoS  results quoted in the first five lines of Table~\ref{tab:summary}.
The PADME measurement, besides being very precise in
the c.m. energy location of the observed excess, has the virtue of coming from
a particle physics experiment. As such it can be taken as completely uncorrelated 
with the nuclear physics results.
As a consequence, the impact of the PADME measurement lies not only in 
increasing the precision of the mass determination, but also in drastically 
reducing the sensitivity of the confidence regions to specific assumptions 
about the treatment of systematic uncertainties.
To illustrate this point, we first construct a covariance matrix $V$ according to the following prescription:
\begin{eqnarray}
\label{eq:V}
V  &=& V^{\rm stat}  + V^{\rm syst} \,,  \\
\label{eq:Vstat}
 V^{\rm stat}_{ij} &=&\delta_{ij} \, (\sigma^{\rm stat}_i)^2 \,,\\
 \label{eq:Vsyst}
 V^{\rm syst}_{ij} &=& {\rm Min}\left[(\sigma^{\rm syst}_i)^2,(\sigma^{\rm syst}_j)^2\right]\,. 
\end{eqnarray}
Using the covariance matrix $V$, we then obtain the chi-squared statistics as  $\chi^2=\Delta^T V^{-1} \Delta$, where $\Delta$ denotes the residual vector representing the differences between observed and expected values.
The prescription for the systematic uncertainties, given 
in  \eq{eq:Vsyst},  aims to be 
as conservative as possible, and is motivated by the following reasoning: 
consider for example the results of the  $^8$Be$^*$ and $^{12}$C$^*$ 
measurements  in the first and fourth row 
in Table~\ref{tab:summary}, with  systematic errors 
$\sigma^{\rm syst}_1 =0.50$ and  $\sigma^{\rm syst}_4=0.20$ respectively. The $^{12}$C$^*$  measurement,  
obtained with an upgraded experimental setup, shows a substantial 
reduction in the systematic uncertainty. The prescription in \eq{eq:Vsyst} amounts to assume that neither the new apparatus, nor the different nucleus used for the measurement,  did  introduce 
 sizeable additional systematic uncertainties uncorrelated with 
$\sigma^{\rm syst}_1$,  and that the residual systematic error is then fully correlated with that of the first  measurement. 
While this procedure allows for a global statistical treatment of the experimental information, it entails approximations.
For example it is very likely 
that the systematic error on the $^8$Be$^*$(GDR) measurement contains components 
that are uncorrelated with  $\sigma^{\rm syst}_1$. 
However, as detailed above, the available information indicates that the uncertainty in the beam-spot position on the target is the dominant source of uncertainty, which is common to all the ATOMKI and VNU-UoS measurements. Therefore, while the assumption of fully correlated systematic errors is rather conservative, it is also well motivated, and we expect that it will not result in a significant overestimation of the overall error.
On the other hand, the most optimistic assumption 
is to consider all systematic errors as uncorrelated, and we have 
performed a second fit to the experimental data adopting this assumption. 
Our results, depicted in Fig.~\ref{fig:Fits} and Fig.~\ref{fig:Chi2}, make apparent 
that after the inclusion of the PADME result, the fits performed with the two 
prescriptions yield very similar results. This demonstrates that the confidence region 
obtained from our fit is robust and reliable, remaining largely unaffected by the 
unknown correlations present in the nuclear physics measurements.


Regarding the particle physics result in the last row 
of Table~\ref{tab:summary}, two intrinsic sources of uncertainty 
on the location of peak of the PADME $e^+e^-$ excess can be readily identified.  The first arises from the beam energy step (0.75\,MeV), and  the second  from the absolute precision in the beam momentum (1.5\,MeV)~\cite{PADME:2024pwg}. These correspond, respectively, to the 0.02\,MeV and 0.05\,MeV uncertainties on $\mX$ listed in the table that we treat as uncorrelated, adding them in quadrature for our fit. 
On the one hand, we expect the c.m. energy uncertainty to dominate over the statistical error from the mass fit.\footnote{\label{foot-pvalues}To check this expectation, we reconstructed an effective $\chi^2$ distribution from the $p$-values shown in Fig.~23 of Ref.~\cite{Bossi:2025ptv}, which yields a statistical uncertainty of $\sim 0.025,\text{MeV}$. As anticipated, this is subdominant compared to the error on the absolute beam momentum. Consequently, including only the systematic uncertainties listed in Table~\ref{tab:summary} while neglecting the statistical error component is well justified.}
On the other hand, since we cannot reliably assign a confidence level to it, the significance 
of our fit should be interpreted with some caution.

In Fig.~\ref{fig:Fits} we plot the results of our fit 
in the two-dimensional plane $\mX-\RBe$. The different contours correspond to the $1\sigma$, $2\sigma$ and  $3\sigma$ confidence level (CL) regions. The left panel corresponds to the measurements given in Table~\ref{tab:summary} with the exclusion of the PADME particle physics measurement. The panel on the right 
includes the PADME result~\cite{LDMA2025,LNFGenSem2025}. In spite of the relatively limited  
statistical evidence of the PADME $e^+e^-$ excess, it is clear that the precision in the determination of the resonant c.m. energy plays a major role in reducing the uncertainty on the mass.
Note that the information on $R_{\rm Be}$ (vertical axis) 
comes from the determinations by the ATOMKI collaboration 
using $^8$Be and by the limits set by the MEG II experiment. For measurements involving other nuclei $N$, as well as for the PADME measurements that solely depend on the $X_{17}$ coupling to electrons $g_{Xe}$, the data are marginalized over  $R_{\rm N}$ and   $g_{Xe}$, respectively.
In Fig.~\ref{fig:Chi2}, we plot the result for the $\chi^2$ fits 
obtained by marginalizing over $\RBe$, i.e. at fixed values of the mass we average out the value of $\RBe$ in the joint distribution. For the nuclear physics only 
case we obtain $\mX =16.78\pm0.12\,$MeV ($1\sigma$ uncertainty)\footnote{A fit to nuclear physics observables including only statistical errors yields $m_{X_{17}}=16.97\pm0.04\MeV$. Another fit to nuclear physics data, using only Be, C and He ATOMKI data, was 
previously performed in Ref.~\cite{Denton:2023gat}, yielding 
$\mX  = 16.85 \pm 0.04 \MeV$.}.
After including the PADME mass determination, the uncertainty gets reduced by more than a factor of two, yielding $\mX = 16.88\pm 0.05\,$MeV. 
Importantly, combining the PADME particle physics result with the results of  nuclear 
physics experiments also resolves the issue of poorly known correlations among 
the systematic errors in the latter measurements.
This is clearly illustrated by comparing the results of the fits performed under the conservative assumption in \eq{eq:Vsyst}  (solid-line contours in Fig.~\ref{fig:Fits} and~\ref{fig:Chi2})
with those obtained under the optimistic assumption of vanishing correlations   
(dashed-line contours in Fig.~\ref{fig:Fits} and~\ref{fig:Chi2}). 
It is evident that, without the PADME data, the unknown correlations would introduce 
a sizeable uncertainty in the overall result
(left panel in Fig.~\ref{fig:Fits} and blue lines in Fig.~\ref{fig:Chi2}).
After including the PADME result, this uncertainty is reduced to a 
negligible level (right panel in Fig.~\ref{fig:Fits} and green lines in Fig.~\ref{fig:Chi2}).

\medskip

\noindent
\section{Outlook and conclusions}  
The possible existence of the X17 boson has been a tantalizing mystery in particle physics for nearly a decade. Initially hinted at by nuclear transition anomalies observed by the ATOMKI Collaboration, the $X_{17}$ has since garnered significant attention, prompting a range of experimental efforts to confirm or refute its existence.
The MEG II experiment recently reported results that place constraints on the existence of such a particle~\cite{MEGII:2024urz}.
In stark contrast, findings from the PADME 
collaboration - an experiment well suited to investigate the $X_{17}$ hypothesis - reveal an intriguing excess with a local significance of approximately $2.5\sigma$. Notably, the analysis was conducted blindly, and the excess emerged precisely within the mass range suggested by the original nuclear physics observations, a coincidence that is, at the very least, remarkable.

In this work, we have presented a comprehensive analysis of experimental data related to the $X_{17}$  boson, encompassing efforts in both nuclear and particle physics. We have  described a methodology for combining the extensive set of nuclear physics results by adopting a dedicated procedure to account for correlations among systematic uncertainties across different measurements. Our approach is based  on the information that can be inferred from the ATOMKI and VNU-UoS publications.
We have shown how the result on the PADME excess significantly sharpens the 
determination of the $X_{17}$ mass. 
We have demonstrated that the combined global result for 
 $m_{X_{17}}$ is robust with respect to specific assumptions about correlations among systematic uncertainties, by performing two sets of fits. 
The first fit was performed under the conservative, yet well-motivated, assumption of positive off-diagonal entries in the correlation matrix (see Eq.~\eqref{eq:Vsyst}). The second fit was carried out under the optimistic (and likely unrealistic) assumption of vanishing correlations. In both cases, upon inclusion of the PADME result, the best-fit value obtained is $m_{X_{17}}=16.88\pm0.05\MeV$.
 We can therefore conclude that incorporating the results of the PADME particle-physics experiment into the dataset effectively resolves the issue of unknown correlations in the nuclear-physics systematic uncertainties, reducing their impact to a negligible level.

We  think that our results can provide robust and reliable prior information for future searches for the $X_{17}$ and related data analyses.

\medskip

\noindent
\textbf{Disclaimer} ---
The analysis and results presented in this study are solely the work of the authors and do not necessarily reflect the views of, nor are they endorsed by, the PADME collaboration.

\medskip


\noindent
\textbf{Acknowledgements} ---
We thank Attila Krasznahorky for providing us with details 
of the ATOMKI analyses and P. Gianotti, M. Raggi, T. Spadaro, P. Valente for discussions.
We thank Luc Darmé for his valuable contributions to shaping the $X_{17}$ search strategy.
F.A.A., G.G.d.C. and E.N. were supported by the INFN ``Iniziativa Specifica" Theoretical Astroparticle Physics (TAsP), with F.A.A. receiving additional support from an INFN Cabibbo Fellowship, call 2022.
G.G.d.C. acknowledges LNF and Sapienza University for hospitality at various stages of this work. The work of E.N. was supported  by the Estonian Research Council grant PRG1884.
We also acknowledge support by  the Estonian Research Council CoE grant TK202 “Foundations of the Universe”, by grants TARISTU24-TK10, TARISTU24-TK3,  by the CERN and ESA Science Consortium of Estonia, grants RVTT3 and RVTT7, and by the COST (European Co-operation in Science and Technology) via the COST Action COSMIC WISPers CA21106. 
The work of C.T. has received funding from the French ANR, under contracts ANR-19-CE31-0016 (`GammaRare') and ANR-23-CE31-0018 (`InvISYble'), that he gratefully acknowledges.

\bibliography{biblio.bib}

\end{document}